\documentclass[12pt,showpacs,showkeys,amsmath,amssymb]{revtex4}
\usepackage{amsmath,amsfonts,amsthm,amscd,amssymb,latexsym}
\usepackage{bm}
\usepackage{dcolumn}
\usepackage{graphicx}
\usepackage{epstopdf}
\usepackage{color}
\usepackage{epsf}
\usepackage{epsfig}
\usepackage{graphicx, epic, eepic, color}

\newcommand{\beq}{\begin{equation}}
\newcommand{\eeq}{\end{equation}}

\begin{document}

\title{Anisotropic Gravitational Collapse and Cosmic Jets}

\author{Donato \surname{Bini}$^{1,2}$}
\email{donato.bini@gmail.com}
\author{Carmen \surname{Chicone}$^{3}$}
\email{chiconec@missouri.edu}
\author{Bahram \surname{Mashhoon}$^{4}$}
\email{mashhoonb@missouri.edu}

\affiliation{
$^1$Istituto per le Applicazioni del Calcolo ``M. Picone'', CNR, I-00185 Rome, Italy\\
$^2$ICRANet, Piazza della Repubblica 10, I-65122 Pescara, Italy\\
$^3$Department of Mathematics and Department of Physics and Astronomy, University of Missouri, Columbia, Missouri 65211, USA\\
$^4$School of Astronomy, Institute for Research in Fundamental
Sciences (IPM), P. O. Box 19395-5531, Tehran, Iran\\
}

\date{\today}

\begin{abstract}
Consider a dynamic general relativistic spacetime in which the proper infinitesimal interval along one spatial coordinate direction decreases monotonically with time, while the corresponding intervals increase along other spatial directions. In a system undergoing such complete anisotropic collapse/expansion, we look for the formation of a cosmic double-jet configuration: free test particles in the ambient medium, relative to the collapsing system, gain energy from the gravitational field and asymptotically line up parallel and antiparallel to the direction of collapse such that their Lorentz factors  approach infinity. A strong burst of electromagnetic radiation is expected to accompany this event if some of the free test particles carry electric charge. Previous work in this direction involved mainly Ricci-flat spacetimes; hence, we concentrate here on inhomogeneous perfect fluid spacetimes. We  briefly explore the possible connection between these theoretical cosmic jets and astrophysical jets. We also discuss other general relativistic scenarios for the formation of cosmic jets. 
\end{abstract}

\pacs{04.20.Cv, 98.58.Fd}
\keywords{Cosmic jets, Astrophysical jets}

\maketitle

\section{Introduction}

Collimated outflows are ubiquitous in astrophysics. In particular, relativistic outflows occur in active galactic nuclei (AGNs) as well as in galactic microquasars~\cite{BPu, Fender, Fender:2014sia}. These relativistic jets consist of magnetically collimated bipolar plasma outflows along the rotation axis of a central collapsed configuration that is believed to be a rotating black hole with an accretion disk. The speed of particles in such jets can sometimes approach the speed of light. 

An intriguing feature of general relativity is that double-jet outflows (``cosmic jets") arise theoretically in certain dynamic anisotropically collapsing configurations. This was first demonstrated via ``gravitomagnetic jets"~\cite{Chicone:2010aa, Chicone:2010hy}. In these theoretically constructed cosmic jets, the speeds of free test particles relative to fiducial static observers asymptotically approach the speed of light. The underlying mechanism for this phenomenon was investigated in Refs.~\cite{Chicone:2010xr, Chicone:2011ie}.  It was later pointed out that a  similar phenomenon occurs in plane wave spacetimes~\cite{Bini:2014esa}. The purpose of the present paper is to study further the characteristic features of these theoretical constructs  that could be related to astrophysical jets. 

Imagine a time-dependent solution of Einstein's gravitational field equations such that the proper infinitesimal distance along one spatial coordinate direction monotonically decreases with time, while at least one other spatial direction expands with time. In this collapse/expansion scenario, we are interested in the asymptotic (i.e., $t \to \infty$) behavior of the solutions of the timelike geodesic equation. The motion of free test particles are referred to a family of observers that are all at rest in space. Previous studies of this problem have indicated the possible formation of  a double-jet pattern along the axis of gravitational collapse in which free test particles move away from the central collapsed configuration such that their speeds relative to the static observers asymptotically approach the speed of light~\cite{Chicone:2010aa, Chicone:2010hy, Chicone:2010xr, Chicone:2011ie, Bini:2014esa}. We speculate that this invariant feature of the cosmic jets comes about because free test particles gain energy from the time-varying gravitational field. 

Consider a stationary gravitational field with a timelike Killing vector field $\xi^\mu$. It is well known that a free test particle with 4-velocity $u^\mu$ has a constant of the motion along its world line given by $\mathbb{E} = -u_\mu\,\xi^\mu$. Time translation invariance implies conservation of total energy and we can assume that the total energy  of the free test particle moving in the stationary gravitational field is constant and proportional to $\mathbb{E}$. In a dynamic spacetime, where the gravitational field is time dependent, we expect that free test particles may exchange energy with the gravitational field in analogy with the pointwise energy exchange that takes place between charged particles moving in an electromagnetic field. However, the pointwise gravitational energy exchange cannot be properly discussed within the framework of general relativity due to the purely local nature of Einstein's principle of equivalence. Therefore, we adopt a different approach that involves the study of timelike geodesics in spacetimes that undergo complete gravitational collapse along a spatial axis. 

The spatially homogeneous Kasner spacetime provides a perfect example for the generation of cosmic jets. The introduction of spatial inhomogeneity, as in the case of the double-Kasner spacetime, can impede the formation of cosmic jets~\cite{Chicone:2011ie}. To gain further insight into the nature of cosmic jets, one can investigate the known time-dependent and spatially inhomogeneous solutions of Einstein's gravitational field equations undergoing asymmetric collapse/expansion. This is a rather daunting task as a perusal of Ref.~\cite{R1} would reveal. We therefore concentrate here on certain simple known solutions of general relativity (GR). The aim of this work is to demonstrate the robustness of the notion of cosmic jets. Previous work in this direction regarding the collapse scenario mainly involved Ricci-flat solutions of GR~\cite{Chicone:2010aa, Chicone:2010hy, Chicone:2010xr, Chicone:2011ie, Bini:2014esa}, whereas here we focus on perfect fluid spacetimes. 

The formation of a cosmic jet is a solitary occurrence. If some of the free test particles in the cosmic jet are electrically charged, then one expects that a strong burst of electromagnetic radiation would result from the formation of a cosmic double-jet configuration. 

Is there a connection between cosmic jets and astrophysical jets? Imagine a spinning cloud of gas and dust that collapses into a disk around its rotation axis.  To treat this system properly within the framework of GR is exceedingly difficult; therefore, we resort to simple exact solutions of GR that represent anisotropic gravitational collapse. Under favorable conditions, we expect the formation of jets of free test particles parallel and antiparallel to the rotation axis. In the ideal case that the thickness of the disk approaches zero, the speed of the double-jet particles would approach the speed of light; otherwise, a milder version of the cosmic jet would occur. If the gravitational collapse results in a rotating black hole surrounded by an accretion disk, particles from the accretion disk can provide a continuous flow to feed the jet via the gravitomagnetic field~\cite{JPGM}. Electromagnetic interactions are then necessary to collimate and maintain the jet.   The structure of the resulting persistent bipolar plasma outflows would depend on the magnetohydrodynamics (MHD) of the astrophysical environment under consideration~\cite{BPu}.

 It is important to emphasize that our gravitational considerations have to do with the \emph{origin} of jets; on the other hand, electromagnetic forces are indispensable to \emph{sustain} the jet.

Sections II, III and IV  demonstrate the collapse scenario for the formation of cosmic jets in certain inhomogeneous perfect fluid spacetimes. In section V, we study the single jet pattern that develops along the direction of propagation of  a plane wave. Section VI demonstrates the limitations of the wave scenario in the case of a nonplanar wave propagation. A discussion of our results is contained in section VII. We use units such that $G= c = 1$, unless specified otherwise. Greek indices run from $0$ to $3$, while Latin indices run from $1$ to $3$. The signature of the spacetime metric is $+2$.

\section{Collapse Scenario: First Example}

Let us consider a perfect fluid spacetime with coordinates $(t, x, y, z)$ that all have dimensions of length. The spacetime metric is given by  
\begin{equation}\label{S1}
ds^2 = - e^{-2x/T_0}\, dt^2 + dx^2+ e^{-(2x+t)/T_0}\,dy^2 + e^{-(2x-t)/T_0}\,dz^2\,,
\end{equation}   
where $T_0$ is a constant length. This Petrov type D solution is a rather special case of the more general  inhomogeneous perfect fluid  solution due to Sintes, Coley and Carot~\cite{Sintes:1997gc}. A discussion of the general case is contained on page 372 of Ref.~\cite{R1}. The general solution, which depends upon a parameter $a$, reduces to Eq.~\eqref{S1} for $a=-1$.  The special solution satisfies the gravitational field equations 
\begin{equation}\label{S2}
R_{\mu \nu} - \frac{1}{2} R\, g_{\mu \nu} + \Lambda\, g_{\mu \nu} = \kappa_0\, T_{\mu \nu}\,,
\end{equation}   
where $\kappa_0 = 8 \pi G/c^4$, $\Lambda$ is the cosmological constant and $T_{\mu \nu}$ is the energy-momentum tensor of a perfect fluid with energy density $\rho$ and pressure $P$,   
\beq \label{S3}
T_{\mu \nu}= (\rho + P)\, U_\mu \,U_\nu + P\, g_{\mu \nu}\,.
\eeq 
Here, $U$, $\rho$ and $P$ are given by 
\beq \label{S4}
 U = e^{x/T_0}\, \partial_t\,, \qquad \kappa_0\,(\rho + P) = -\frac{1}{2\,T_0^2}\,e^{2x/T_0}\,, \qquad \kappa_0\,P-\Lambda= \frac{12-e^{2x/T_0}}{4\,T_0^2}\,.
\eeq  
The Kretschmann scalar $K$,
\beq \label{S5}
K := R_{\,\alpha \beta \gamma \delta} R^{\,\alpha \beta \gamma \delta}\,,
\eeq 
for this solution is given by
\beq \label{S5a}
K = \frac{96 - 8\, e^{2x/T_0} + 3 \,e^{4x/T_0}}{4\,T_0^2}\,.
\eeq 
It follows from these results that there is a curvature singularity at $x=\infty$; moreover, the density and pressure are unphysical and do not depend upon time in this collapsing configuration. However,  this circumstance appears to have little direct bearing on the cosmic jet that develops in this solution. We note that $\sqrt{-g} = \exp(-3x/T_0)$ vanishes as $x \to \infty$. There are three Killing vector fields in this spacetime, namely, 
\beq \label{S6}
 2\,T_0\, \partial_t + y\, \partial_y - z\, \partial_z\,, \qquad \partial_y\,, \qquad \partial_z\,.
\eeq    
We now turn to the solution of the timelike geodesic equation. 

As is well known, the geodesic equation can be obtained via the Euler-Lagrange equation associated with a Lagrangian that is given, up to a proportionality constant, by $(ds/d\tau)^2$, where $\tau$ is the proper time along the geodesic world line with unit tangent vector $u^\mu = dx^\mu/d\tau$. From the projection of $u^\mu$ upon the Killing vector fields $\partial_y$ and $\partial_z$, we find 
\begin{equation}\label{S7}
\frac{dy}{d\tau} = C_y\, e^{(2x+t)/T_0}\,, \qquad \frac{dz}{d\tau} = C_z\, e^{(2x-t)/T_0}\,,
\end{equation} 
where $C_y$ and $C_z$ are constants of the motion. Moreover,  $u_{\mu}\,u^{\mu} =-1$ implies that
\begin{equation}\label{S8}
e^{-2x/T_0}\,\left(\frac{dt}{d\tau}\right)^2 -(1+ \dot{x}^2) = C_y^2\, e^{(2x+t)/T_0} + C_z^2\, e^{(2x-t)/T_0}\,,
\end{equation}  
where $\dot{x} := dx/d\tau$. Using Eq.~\eqref{S7}, the remaining equations for the motion of a free test particle can be expressed as   
\begin{equation}\label{S9}
\frac{d}{d\tau}\left(e^{-2x/T_0}\,\frac{dt}{d\tau}\right) =\frac{1}{2T_0}\left[C_y^2\, e^{(2x+t)/T_0} - C_z^2\, e^{(2x-t)/T_0}\right]\,
\end{equation}
and
\begin{equation}\label{S10}
\ddot{x} =\frac{1}{T_0}\,e^{-2x/T_0}\,\left(\frac{dt}{d\tau}\right)^2 - \frac{1}{T_0}\left[C_y^2\, e^{(2x+t)/T_0} + C_z^2\, e^{(2x-t)/T_0}\right]\,.
\end{equation}  

To solve these equations, we note that Eqs.~\eqref{S8} and~\eqref{S10} can be combined to get 
\begin{equation}\label{S11}
T_0\,\ddot{x} = 1+ \dot{x}^2\,.
\end{equation}  
Integrating this equation once, we find
\begin{equation}\label{S11a}
\dot{x} = \tan\left(\frac{\tau + B}{T_0}\right)\,,
\end{equation}  
which, upon further integration, implies that 
\begin{equation}\label{S11b}
e^{-x/T_0} = A\,\cos \left(\frac{\tau + B}{T_0}\right)\,,
\end{equation}  
where $A \ne 0$ and $B$ are constants of integration. Suppose that at the initial proper time $\tau=0$, $(t, x, y, z) = (t_i, x_i, y_i, z_i)$. To avoid the curvature singularity ($x=\infty$) at this initial event, we must assume that $\cos(B/T_0) \ne 0$;  to simplify matters, we  henceforth choose  $B$ such that $0 \le B/T_0 < \pi/2$. Indeed, with the initial conditions that at $\tau=0$, $x=x_i$ and $\dot{x}=\dot{x}_i$ are finite, the solution of Eq.~\eqref{S11} is given by 
\begin{equation}\label{S12}
e^{-(x-x_i)/T_0} = \cos (\tau/T_0) - \dot{x}_i \,\sin (\tau/T_0)\,.
\end{equation}
Next, we define   
\begin{equation}\label{S13}
e^{-2x/T_0}\,\frac{dt}{d\tau} := F\,,
\end{equation} 
and write Eq.~\eqref{S9} as 
\begin{equation}\label{S14}
2F\,\frac{dF}{dt} =\frac{1}{T_0}\left[C_y^2\, e^{t/T_0} - C_z^2\, e^{-t/T_0}\right]\,.
\end{equation}
Integrating this equation we find
\begin{equation}\label{S15}
F^2 =C_y^2\, e^{t/T_0} + C_z^2\, e^{-t/T_0} +C\,,
\end{equation}
where $C$ is an integration constant. 
We assume that the temporal coordinate increases monotonically with proper time along the timelike geodesic; hence, it follows from Eq.~\eqref{S13} that we must choose $F(t)$ to be positive, namely,  
\begin{equation}\label{S16}
F =  \left[C_y^2\, e^{t/T_0} + C_z^2\, e^{-t/T_0} + C\right]^{1/2}\,,
\end{equation}
where positive square roots are considered throughout. From Eqs.~\eqref{S11b}, ~\eqref{S13} and~\eqref{S16}, we find
\begin{equation}\label{S17}
\frac{A^2}{T_0} \int_{t_i}^t \frac{dt'}{F(t')} =   \tan\left(\frac{\tau + B}{T_0}\right) -  \tan\left(\frac{B}{T_0}\right)\,.
\end{equation}
As $t \to \infty$, the left-hand side of this relation is finite, which means that $\tau \to \tau_f$ such that $\tau_f +B <\pi\,T_0/2$. It follows that the domain of variation of $x$ is finite throughout the geodesic motion and free test particles stay away from the curvature singularity at $x=\infty$. 

Let us now analyze the motion of free test particles in this spacetime with respect to comoving observers that are at rest in space with 4-velocity vector $U = e_{\hat 0}$. The natural orthonormal tetrad frame of such observers is given by
\begin{eqnarray}\label{S18}
e_{\hat 0} = e^{x/T_0}\,\partial_t\,,\qquad
e_{\hat 1} = \partial_x\,, \qquad
e_{\hat 2} = e^{(2x+t)/(2T_0)}\,\partial_y\,, \qquad
e_{\hat 3} = e^{(2x-t)/(2T_0)}\,\partial_z\,.
\end{eqnarray}
The 4-velocity vector of the free test particles relative to the static observers can be expressed as
\begin{equation}\label{S19}
u_{\hat \alpha} =  u_\mu\,e^{\mu}{}_{\hat \alpha} := \Gamma\,(-1, V_x, V_y, V_z)\,,
\end{equation} 
where the geodesic particle's Lorentz factor is given by
\begin{equation}\label{S20}
\Gamma = e^{x/T_0}\,F(t)\,
\end{equation} 
and 
\begin{equation}\label{S21}
V_x = \frac{1}{\Gamma}\,\tan\left(\frac{\tau + B}{T_0}\right)\,, \qquad  V_y= C_y\, \frac{e^{t/(2T_0)}}{F(t)}\,, \qquad V_z = C_z\, \frac{e^{-t/(2T_0)}}{F(t)}\,.
\end{equation} 
As $t \to \infty$, the range of $x$ is finite and $F(t) \sim |C_y|\, \exp(\frac{1}{2}t/T_0)$; hence $\Gamma \to \infty$. Moreover, in this limit, $(V_x, V_y, V_z) \to (0, \pm 1, 0)$, resulting in a double-jet configuration in a region free of curvature singularities along the $y$ direction, which is the collapse axis in this case.

\section{Collapse Scenario: Second Example}

Next, let us consider the spacetime metric given by 
\begin{equation}\label{M1}
ds^2 = - e^{-2x/T_0}\, dt^2 + dx^2+ e^{-2x/T_0}\, \tilde{t}^{\,1+b}\,dy^2 + e^{-2x/T_0}\, \tilde{t}^{\,1-b}\,dz^2\,,
\end{equation}   
where $ \tilde{t} = t/T_0$, $T_0$ is a constant length as before and $b$ is a dimensionless  constant. This solution is of Petrov type $D$ and is a particular case of the  general  inhomogeneous perfect fluid solution due to Mars and Wolf~\cite{Mars:2002pn}, see Appendix A. As before, it follows from the gravitational field equations that $U$, $\rho$ and $P$ are given by 
\beq \label{M2}
 U = e^{x/T_0} \,\partial_t\,, \qquad \kappa_0\,(\rho + P) = -\frac{b^2-1}{2t^2}\,e^{2x/T_0}\,, \qquad \kappa_0\,P-\Lambda= \frac{3}{T_0^2}-\frac{b^2-1}{4t^2}\,e^{2x/T_0}\,.
\eeq  
The Kretschmann scalar for this solution is 
\beq \label{M3}
 K = \frac{96\,t^4 - 8\,(b^2-1)t^2T_0^2\, e^{2x/T_0} + 3 \,T_0^4 (b^2-1)^2\,e^{4x/T_0}}{4\,t^4T_0^4}\,.
\eeq   
It follows from these results that there are curvature singularities at $t=0$ and $x=\infty$, where $\sqrt{-g} = (t/T_0)\,\exp(-3x/T_0)$ vanishes as well. The perfect fluid source becomes unphysical for $b^2 > 1$.     As before, there are three Killing vector fields
\beq \label{M4}
 2\,t\, \partial_t + 2\,T_0\, \partial_x + (1-b)\,y \,\partial_y + (1+b)\, z\, \partial_z\,, \qquad \partial_y\,, \qquad \partial_z\,.
\eeq 

The physical characteristics of this solution depend upon $b^2$. Thus with no loss in generality we assume $b>0$. Furthermore, to ensure that we have an anisotropically collapsing configuration, we must assume
\beq \label{M4a}
b>1\,,
\eeq 
in the rest of this section. This means that, just as in section II, the perfect fluid source here is essentially unphysical.
   
The similarity between this spacetime and the one in the previous section implies that the equations of motion of free test particles are closely related. In fact, there are constants of the motion $C_2$ and $C_3$ due to the existence of the 
Killing vector fields $\partial_y$ and $\partial_z$ such that 
\begin{equation}\label{M5}
\frac{dy}{d\tau} = C_2\, e^{2x/T_0}\, \tilde{t}^{\,-(1+b)}\,, \qquad \frac{dz}{d\tau} = C_3\, e^{2x/T_0}\, \tilde{t}^{\,-(1-b)}\,.
\end{equation}  
Furthermore, it follows from $u_{\mu}\,u^{\mu} =-1$ that
\begin{equation}\label{M6}
e^{-2x/T_0}\,\left(\frac{dt}{d\tau}\right)^2 -(1+ \dot{x}^2) = e^{2x/T_0}\,\left[C_2^2\,  \tilde{t}^{\,-(1+b)} + C_3^2\, \tilde{t}^{\,-(1-b)} \right]\,.
\end{equation} 
Moreover, we have the additional equations of motion
\begin{equation}\label{M7}
\frac{d}{d\tau}\left(e^{-2x/T_0}\,\frac{dt}{d\tau}\right) = -\frac{T_0\,e^{2x/T_0}}{2t^2}\left[C_2^2\, (1+b)\, \tilde{t}^{\,-b} + C_3^2\, (1-b)\, \tilde{t}^{\,b}\right]\,
\end{equation}
and
\begin{equation}\label{M8}
\ddot{x} =\frac{1}{T_0}\,e^{-2x/T_0}\,\left(\frac{dt}{d\tau}\right)^2 - \frac{1}{T_0}\,e^{2x/T_0}\,\left[C_2^2\,  \tilde{t}^{\,-(1+b)} + C_3^2\,  \tilde{t}^{\,-(1-b)} \right]\,.
\end{equation} 

To solve these differential equations, we assume that at some initial proper time $\tau=0$, $(t, x, y, z) = (t_i, x_i, y_i, z_i)$, where $t_i >0$ in order to avoid the $t=0$ curvature singularity at this initial event. It follows from Eqs.~\eqref{M6} and~\eqref{M8} that the $x$ coordinate of a free particle satisfies exactly the same equation as  Eq.~\eqref{S11}. To avoid the $x=\infty$ curvature singularity at the initial proper time $\tau=0$, we have as before
\begin{equation}\label{M9}
\dot{x} = \tan\left(\frac{\tau + B_0}{T_0}\right)\,,\qquad e^{-x/T_0} = A_0\,\cos \left(\frac{\tau + B_0}{T_0}\right)\,,
\end{equation}  
where $A_0 \ne 0$ and $B_0$ are integration constants and $0 \le B_0/T_0 < \pi/2$. Let us now define the function $\Phi$,
\begin{equation}\label{M10}
\Phi := e^{-2x/T_0}\,\frac{dt}{d\tau}\,;
\end{equation} 
then, as before, Eq.~\eqref{M7} can be integrated and the solution for $dt/d\tau >0$ is 
\begin{equation}\label{M11}
\Phi (t) = \left[C_2^2\,  \tilde{t}^{\,-(1+b)} + C_3^2\,  \tilde{t}^{\,-(1-b)} + C_0 \right]^{1/2}\,.
\end{equation} 
Using Eqs.~\eqref{M9}--\eqref{M11}, it is possible to express $t$ as a function of $\tau$, namely, 
\begin{equation}\label{M12}
\frac{A_0^2}{T_0} \int_{t_i}^t \frac{dt'}{\Phi(t')} =   \tan\left(\frac{\tau + B_0}{T_0}\right) -  \tan\left(\frac{B_0}{T_0}\right)\,,
\end{equation}
where $t_i>0$ by assumption. As $t\to \infty$, the left-hand side of this equation diverges for $1 < b \le 3$, which means that $\tau_f +B_0 = \pi \, T_0/2$ and $x_f = \infty$, while for $b>3$, the left-hand side remains finite, $\tau_f +B_0 < \pi \, T_0/2$, and the range of $x$ is finite.

To describe the motion of free test particles relative to comoving observers, we note that the natural orthonormal frame of static observers is given by
\begin{eqnarray}\label{M13}
e_{\hat 0} = e^{x/T_0}\,\partial_t \,, \qquad
e_{\hat 1} = \partial_x\,, \qquad
e_{\hat 2} = e^{x/T_0}\, \tilde{t}^{\,-(1+b)/2}\,\partial_y\,, \qquad
e_{\hat 3} = e^{x/T_0}\, \tilde{t}^{\,-(1-b)/2}\,\partial_z\,.
\end{eqnarray}
The projection of the 4-velocity vector of the free test particles $u_\mu$ onto the tetrad frame of the static observers $e^{\mu}{}_{\hat \alpha}$ is given, as before,  
by $u_{\hat \alpha} =  u_\mu\,e^{\mu}{}_{\hat \alpha} := \Gamma\,(-1, V_x, V_y, V_z)$. 
Hence, the relative Lorentz factor is given by $\Gamma = \exp{(x/T_0)}\,\Phi(t)$, where as $t \to \infty$, 
$\Phi$ has the asymptotic behavior given by
$\Phi \sim  |C_3|\, \tilde{t}^{\,-(1-b)/2}$. It follows from $b>1$  that as $t \to \infty$, $\Phi \to \infty$. On the other hand, $\exp(x/T_0)$ diverges asymptotically for $1< b \le 3$, but is finite for $b>3$. In either case,  $\Gamma$ diverges asymptotically. Specifically, one can show that in the spacetime under consideration,
\begin{equation}\label{M14}
V_x = \frac{A_0}{\Phi(t)}\,\sin\left(\frac{\tau + B_0}{T_0}\right)\,, \qquad  V_y= \frac{C_2}{\Phi(t)}\,   \tilde{t}^{\,-(1+b)/2}\,, \qquad V_z = \frac{C_3}{\Phi(t)}\,   \tilde{t}^{\,-(1-b)/2}\,.
\end{equation} 
As $t \to \infty$,
\begin{equation}\label{M15}
(V_x, V_y, V_z) \to (0, 0, \pm1),
\end{equation}
which is again a double-jet configuration in the direction of collapse. The cosmic jet occurs in a region free of spacetime singularities for $b>3$, while for  $1< b \le 3$, the free test particles end up at the curvature singularity $x=\infty$.

\section{Collapse Scenario: Third Example}

Finally, we consider a solution due to Mars and Wolf~\cite{Mars:2002pn} that can be expressed as
\begin{equation}\label{W1}
ds^2 = -\tilde{x}^{2Q}\, dt^2 + \tilde{t}^{\,2p_1}\,dx^2+ \tilde{x}^{2Q}\,\tilde{t}^{\,2p_2}\,dy^2 + \tilde{x}^{2Q}\,\tilde{t}^{\,2p_3}\,dz^2\,,
\end{equation}  
where
\begin{equation}\label{W2}
\tilde{t} := \frac{t}{T_0}\,, \qquad \tilde{x} := \frac{x}{L_0}\,.
\end{equation} 
Here, $T_0$ and $L_0$ are constant lengths and we assume that in general $t > 0$ and $x > 0$. Moreover, 
\begin{equation}\label{W3}
p_1 =  \frac{\alpha}{\alpha+2}\,, \qquad p_2= \frac{1-\beta}{\alpha+2}\,, \qquad p_3=\frac{1+\beta}{\alpha+2}\,, \qquad Q = \frac{2\alpha+1-\beta^2}{\alpha^2}\,,
\end{equation}  
where $\alpha$ and $\beta$ are constant parameters such that $\alpha \ne 0, -2$. For $Q=0$, we recover the  Kasner   solution, while for $Q\ne 0$, we have a solution that is conformally related to a Kasner-like solution, see Appendix A. 

The source of the Mars-Wolf solution is a perfect fluid that is in motion along the $x$ direction. That is, let us consider observers that are at rest in space. The natural tetrad frame $e^{\mu}{}_{\hat \alpha}$ of these static observers is given by 
\begin{eqnarray}\label{W4}
e_{\hat 0} = \tilde{x}^{-Q}\,\partial_t\,, \qquad
e_{\hat 1} = \tilde{t}^{\,-p_1}\,\partial_x\,,\qquad
e_{\hat 2} =  \tilde{x}^{-Q}\,  \tilde{t}^{\,-p_2}\,\partial_y\,, \qquad
e_{\hat 3} =  \tilde{x}^{-Q}\,  \tilde{t}^{\,-p_3}\,\partial_z\,.
\end{eqnarray}
With respect to this tetrad frame, the 4-velocity of the perfect fluid can be written as 
\beq\label{W5}
U=\gamma_0 (e_{\hat 0} + \nu_0\, e_{\hat 1})\,,\qquad \gamma_0 =(1-\nu_0^2)^{-1/2}\,.
\eeq
 We can now determine the energy density and the pressure of the fluid from the gravitational field equations. The results are 
\begin{equation}\label{W5}
\kappa_0 (\rho + P) = \frac{2Qp_1^2}{t^2}\, \tilde{x}^{-2Q}\,(1-\nu_0^2)\,, 
\end{equation} 
\begin{equation}\label{W6}
\kappa_0\, P -\Lambda =  \frac{Qp_1^2}{t^2}\, \tilde{x}^{-2Q} -  \frac{Q(2-3Q)}{x^2}\, \tilde{t}^{\,-2p_1}\,.
\end{equation} 
Moreover, the fluid velocity is given by 
\begin{equation}\label{W7}
\nu_0 = - \frac{t}{p_1x}\, \tilde{x}^{\,Q}\, \tilde{t}^{\,-p_1}\,,
\end{equation} 
which should satisfy the requirement that 
\begin{equation}\label{W7a}
-1 < \nu_0 < 1\,.
\end{equation}

Let us now assume that the cosmological constant vanishes. With $\Lambda = 0$, $\rho$ and $P$ are given by
\begin{equation}\label{W7b}
\kappa_0 \,\rho = \frac{Qp_1^2}{t^2}\, \tilde{x}^{-2Q} - \frac{3Q^2}{x^2}\, \tilde{t}^{\,-2p_1}\,, 
\end{equation}
\begin{equation}\label{W7c}
\kappa_0 \,P = \frac{Qp_1^2}{t^2}\, \tilde{x}^{-2Q} - \frac{Q(2-3Q)}{x^2}\, \tilde{t}^{\,-2p_1}\,, 
\end{equation}
so that
\begin{equation}\label{W7d}
\kappa_0 \,(\rho-P) =  \frac{2Q(1-3Q)}{x^2}\, \tilde{t}^{\,-2p_1}\,. 
\end{equation} 
It is clear that if $Q < 0$, then $\rho < 0$  and $\rho < P$. On the other hand, if $Q>0$ and $\rho > 0$, then $3Q\,\nu_0^2 < 1$; moreover,  $P> \rho >0$ for $Q>1/3$, while for $Q<1/3$, $P< \rho$. For $Q=1/3$, we have $P = \rho$. Therefore, we must have $0 < Q \le 1/3$ to ensure that reasonable energy conditions can be satisfied for the perfect fluid source under consideration here; furthermore, we must maintain condition~\eqref{W7a} as well. 

The Kretschmann scalar for the Mars-Wolf solution simplifies and is given by 
\beq \label{W8}
 K = 4\,\frac{3Q^2p_1^4-4p_1p_2p_3}{t^4}\, \tilde{x}^{-4Q} + 8\, \frac{p_1^2 Q^2(Q-2)}{t^2x^2}\, \tilde{t}^{\,-2p_1}\,\tilde{x}^{-2Q} + 12\,\frac{Q^2(1-2Q+2Q^2)}{x^4}\, \tilde{t}^{\,-4p_1}\,.
\eeq   
It follows that in general $t=0$ and $x=0$ are curvature singularities of the Mars-Wolf spacetime. Let us note that $\sqrt{-g}=\tilde{t}\,\tilde{x}^{\,3Q}$. Moreover, there are two spacelike Killing vector fields in this spacetime,
\beq \label{W9}
\partial_y\,, \qquad \partial_z\,,
\eeq 
and a homothetic vector field given by
\beq \label{W10}
\frac{1-Q}{1-p_1}\, t\, \partial_t + x\, \partial_x + \frac{1-Q}{1-p_1}\, (1-p_2)\,y\, \partial_y + \frac{1-Q}{1-p_1}\,(1-p_3)\, z\, \partial_z\,,
\eeq
which reduces to $x\, \partial_x$ for $Q = 1$.

\subsection{Timelike Geodesics}

The motion of free test particles in the Mars-Wolf spacetime involves two constants of the motion $\mathcal{C}_y$ and $\mathcal{C}_z$ that can be obtained from the projection  of $u^\mu$, the
4-velocity vector of a geodesic particle, upon the Killing vectors  $\partial_y$ and $\partial_z$, respectively. It is then useful to define 
\begin{equation}\label{W11}
\mathcal{F}(t) = \mathcal{C}_y^2 ~ \tilde{t}^{\,-2p_2} + \mathcal{C}_z^2~\tilde{t}^{\,-2p_3}\,,
\end{equation}
since $u_\mu\,u^\mu = -1$ can be written as
\begin{equation}\label{W11}
\tilde{x}^{\,2Q}\,\left(\frac{dt}{d\tau}\right)^2 = 1+ \tilde{t}^{\,2p_1}\,\dot{x}^2 + \tilde{x}^{-2Q}\, \mathcal{F}(t)\,.
\end{equation}  
Furthermore, let us introduce $\mathcal{W}$,
\begin{equation}\label{W12}
\mathcal{W} := \tilde{t}^{\,2p_1}\,\dot{x}\,.
\end{equation}  
Then, the equations of motion of free test particles are given by
\begin{eqnarray}\label{W13}
\frac{dt}{d\tau} &=& \tilde{x}^{-Q}\, \Gamma\,, \nonumber\\
\frac{dx}{d\tau} &=& \tilde{t}^{\,-2p_1}\, \mathcal{W}\,, \nonumber\\ 
\frac{d\mathcal{W}}{d\tau} &=& -\frac{Q}{x}\,\left(1+\tilde{t}^{\,-2p_1}\, \mathcal{W}^2\right)\,, \nonumber\\
\frac{dy}{d\tau} &=& \mathcal{C}_y\, \tilde{x}^{-2Q}\, \tilde{t}^{\,-2p_2}\,, \nonumber\\
\frac{dz}{d\tau} &=& \mathcal{C}_z\, \tilde{x}^{-2Q}\, \tilde{t}^{\,-2p_3}\,,
\end{eqnarray}
where
\begin{equation}\label{W14}
\Gamma =\left[1+ \tilde{t}^{\,-2p_1}\, \mathcal{W}^2 + \tilde{x}^{-2Q}\,\mathcal{F}(t)\right]^{1/2}\,.
\end{equation}

It has not been possible to find analytic solutions of these equations. We must therefore numerically integrate these equations with the  initial conditions that at $\tau =0$, we have the following \emph{initial} values: $t(0), x(0), \mathcal{W}(0), y(0)$ and $z(0)$. We integrate \emph{forward} in proper time $\tau$ towards $t=\infty$ and \emph{backward} towards the $t = 0$ singularity. As before, we wish to express the motion of free test particles relative to observers that are at rest in the Mars-Wolf spacetime. Using the tetrad frame~\eqref{W4} and 
$u_{\hat \alpha} =  u_\mu\,e^{\mu}{}_{\hat \alpha} := \Gamma\,(-1, V_x, V_y, V_z)$, we find 
\begin{equation}\label{W15}
V_x= \tilde{t}^{\,-p_1} \, \frac{\mathcal{W}}{\Gamma}\,, 
\end{equation}
\begin{equation}\label{W16}
V_y= \frac{\mathcal{C}_y}{\Gamma}\,\tilde{x}^{-Q} \tilde{t}^{\,-p_2}\,, 
\end{equation}
\begin{equation}\label{W17}
V_z= \frac{\mathcal{C}_z}{\Gamma}\,\tilde{x}^{-Q} \tilde{t}^{\,-p_3}\,. 
\end{equation}

It is interesting to note here that for some metric parameters, such as $(\alpha, \beta) = (4, \pm3)$, $Q = 0$; then, $\mathcal{W}$ is constant, $\mathcal{W} = \mathcal{C}_x$, and our treatment reduces  to the determination of cosmic jets in Kasner spacetime~\cite{Chicone:2010xr}. In fact, one can compare and contrast our $Q \ne 0$ treatment with the Kasner case to determine the impact of inhomogeneity on the formation of cosmic jets.  

For $Q \ne 0$, the behavior of the timelike geodesics along the $x$ direction crucially depends upon whether $Q>0$ or $Q<0$. In fact, inspection of the geodesic equations of motion~\eqref{W13} reveals that upon forward integration in the   $Q>0$ case, an initial positive value of coordinate $x$ monotonically decreases to the curvature singularity at $x=0$, while in the $Q<0$ case, $x$ would monotonically increase away from the singularity.  Furthermore, $Q<0$ automatically implies that $\rho <0$ and the perfect fluid source is unphysical, while in the  $Q>0$ case, we can only ensure that the world lines of the free particles \emph{initially} pass through a perfect fluid medium with reasonable physical properties.   

Indeed, unlike the simpler solutions treated in sections II and III, we have some control here over the nature of the source. That is, for $Q>0$, we can ensure that the initial conditions are such that at the beginning the timelike geodesics go through the perfect fluid where  $\rho \ge P > 0$ and $-1<\nu_0 <1$. As an example, let $(\alpha, \beta) = (3, -2)$, so that $Q=1/3$ and $p_1=  p_2 = 3/5, p_3 =-1/5$; hence, it follows from Eq.~\eqref{W7d} that $\rho = P$ in this case.  To ensure that $\rho >0$ and hence 
$\nu_0^2 <1$ initially at $\tau =0$, we let $t(0)= 1, x(0) =10^3, \mathcal{W}(0) =0, y(0)=0$ and  $z(0)=0$.  Moreover, in the geodesic equations~\eqref{W13}, we assume $\mathcal{C}_y=\mathcal{C}_z=1$ and $T_0= L_0 = 1$ unit of length, so that $\tilde{t}=t$ and $\tilde{x}=x$. We recall that there are curvature singularities at $x=0$ and $t=0$. In this case, collapse takes place along the $z$ axis but  forward  numerical integration runs into the $x=0$ curvature singularity at $\tau \approx 3.428 \times 10^6$, while the formation of a cosmic jet along the $z$ axis is taking place,  see Figures 1 and 2. We note that in this case, the nature of the perfect fluid along the geodesic world line is reasonable up to about $\tau = 2\times 10^5$, but after that $\rho < 0$ and perfect fluid motion is superluminal. The 
presence of spatial inhomogeneity in the Mars-Wolf solution results in the existence of the 
curvature singularity at $x = 0$, which for $Q>0$ generally intervenes in the formation of  cosmic jets. 

Let us now consider \emph{backward} integration of geodesic equations of motion from $\tau =0$ with the same initial conditions as before towards the other curvature singularity at $t = 0$. Along the geodesic world line, the perfect fluid source is completely reasonable, a cosmic jet develops in the $(x, y)$ plane that is mostly in the $y$ direction with a deviation angle of $0.03$ radians towards the $x$ direction and
with $\Gamma \to \infty$  as we reach the $t = 0$ singularity at $\tau \approx -9.612$. 

\begin{figure}
\begin{center}
\includegraphics[width=10 cm]{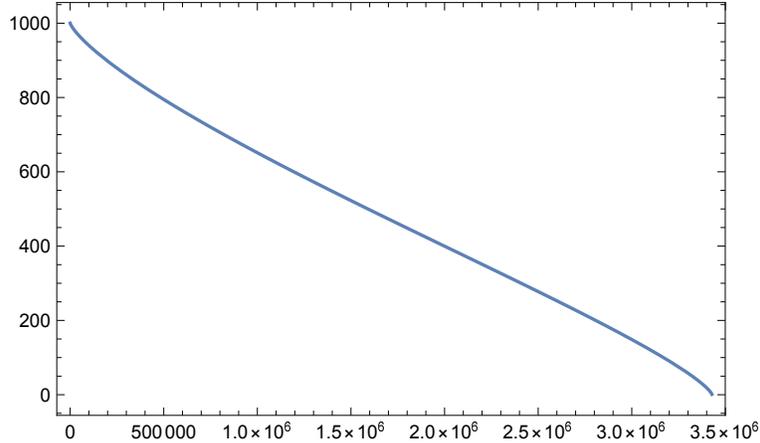}
\end{center}
\caption{Graph of  $x$  versus proper time $\tau$ for the parameter values $(\alpha, \beta) = (3,-2)$ and $Q=1/3$. At $\tau = 0$, we have $x(0) =  10^3, t(0) = 1, \mathcal{W}(0) = 0$ and $y(0) = z(0) = 0$. Furthermore, we assume $\mathcal{C}_y = \mathcal{C}_z = 1$.}
\label{fig1} 
\end{figure}

\begin{figure}
\begin{center}
\includegraphics[width=5 cm]{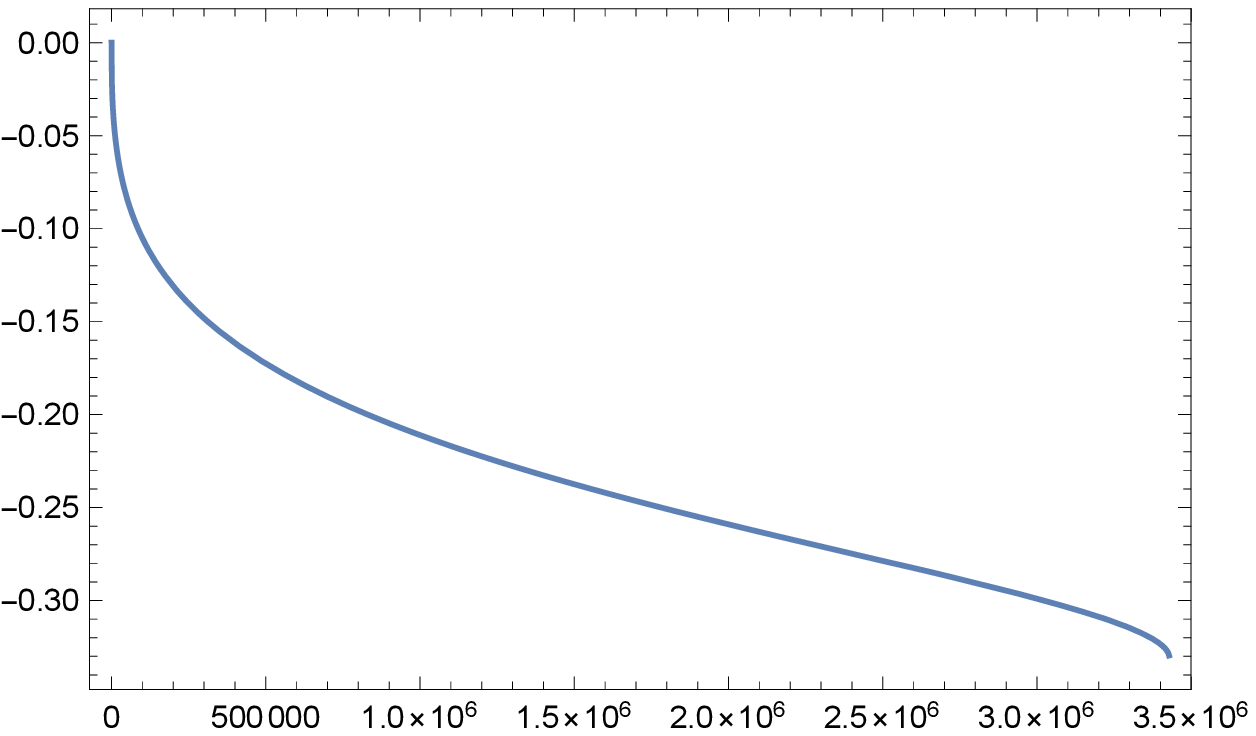}\quad \includegraphics[width=5 cm]{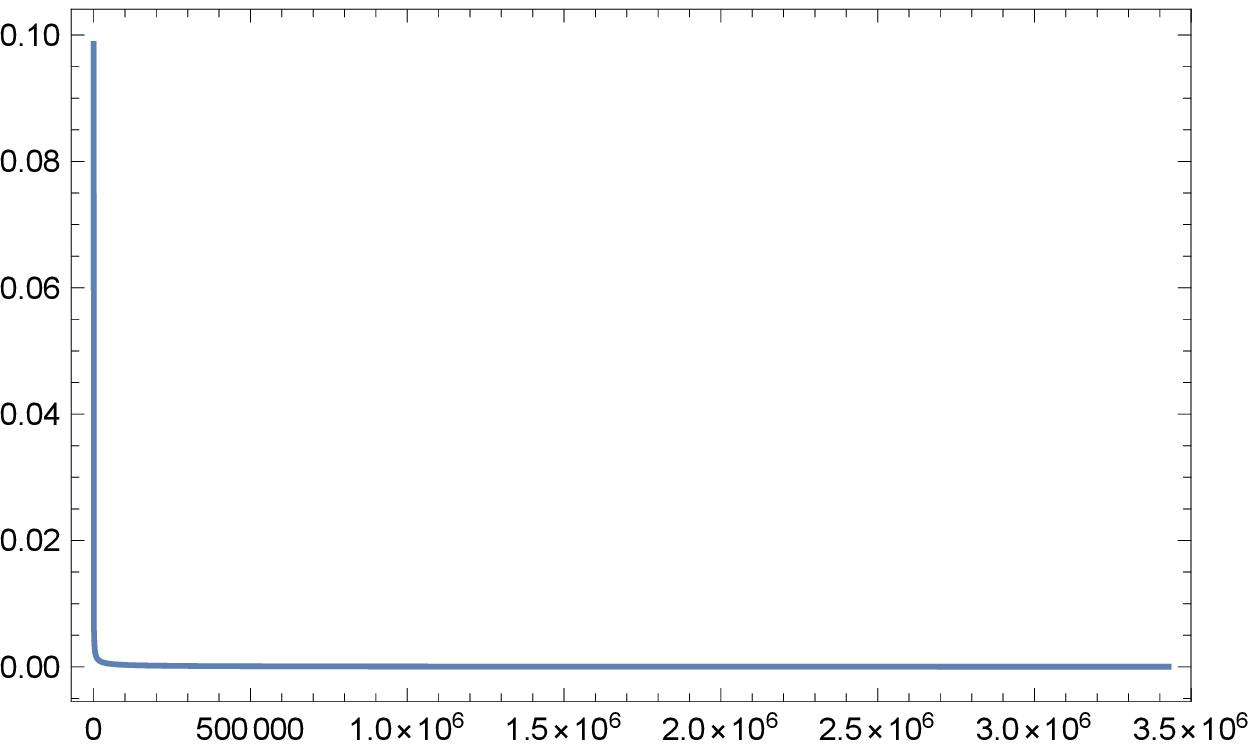}\quad \includegraphics[width=5 cm]{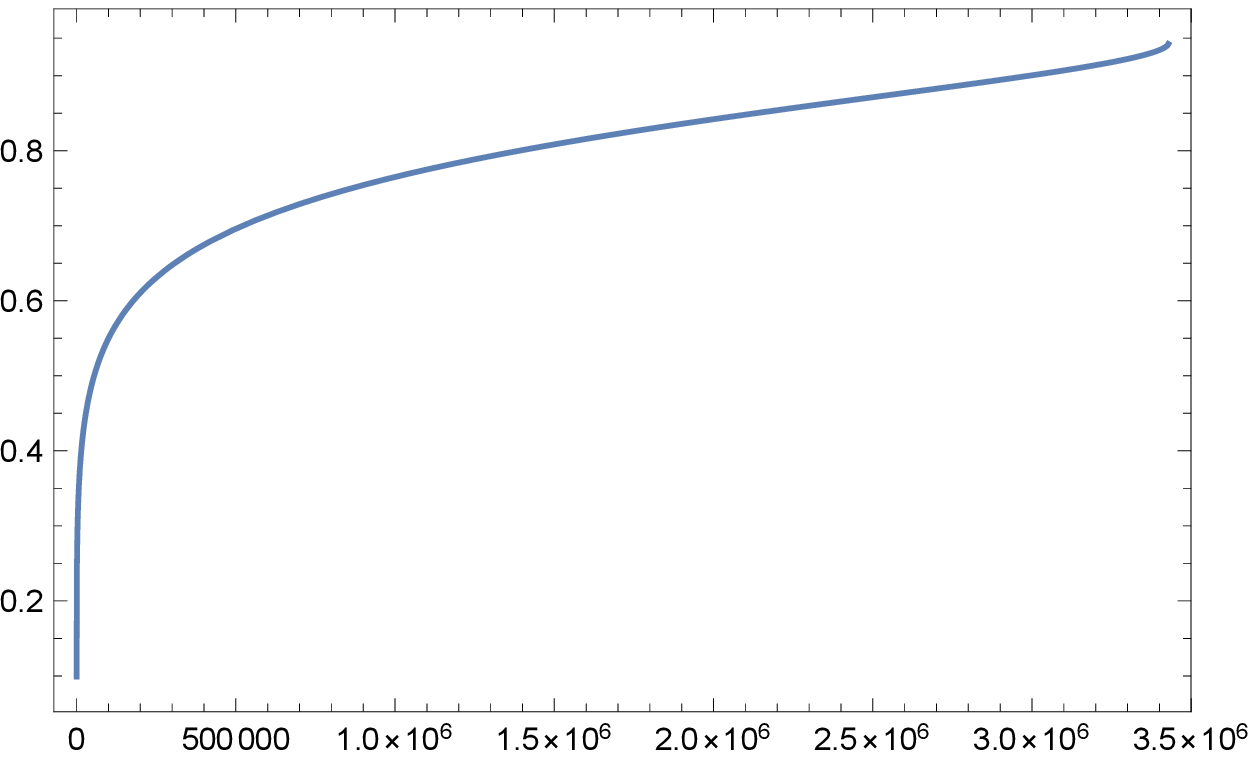}
\end{center}
\caption{ The figure depicts,  from left to right,  graphs of $V_x$, $V_y$ and $V_z$  versus $\tau$  for the parameter values $(\alpha, \beta) = (3,-2)$ and $Q=1/3$. We assume that at $\tau =0, t(0) = 1, x(0) = 10^3, \mathcal{W}(0) = 0$ and $y(0) = z(0) = 0$. As before, $\mathcal{C}_y = \mathcal{C}_z = 1$.}
\label{fig2} 
\end{figure}

For the sake of completeness, we consider next an example where $Q< 0$. That is, let $(\alpha, \beta) = (-1, -1)$, so that $p_1= -1, p_2 = 2, p_3 =0$ and $Q = -2$. It follows that $\rho < 0$ and the fluid source is unphysical. We assume $t(0) = 1/2, x(0) =1/2, \mathcal{W}(0) = 0, y(0) =0, z(0) = 0$ and  $\mathcal{C}_y=\mathcal{C}_z=1$. In this case, collapse takes place along the $x$ direction and a cosmic jet develops along the axis of collapse as shown in Figure 3.  

In the particular example under consideration with $Q=-2$, the range of the spacetime coordinates $(t, x, y, z)$ is unrestricted. Moreover,  Eq.~\eqref{W13} remains invariant under the transformations $x \mapsto -x$ and $\mathcal{W} \mapsto -\mathcal{W}$. The result is that $V_x \mapsto -V_x$, but $V_y$ and $V_z$ remain the same. This means that if we start with  $x(0) = -1/2$, forward integration results in the cosmic jet antiparallel to the $x$ direction.  For an initial random distribution of free test particles, a double-jet configuration develops asymptotically along the $x$ direction with $\Gamma \to \infty$ and
\begin{equation}\label{W18}
 (V_x, V_y, V_z) \to (\pm 1, 0, 0)\,.
\end{equation}  

It is interesting to consider \emph{backward} integration in proper time towards $t = 0$ with our original initial conditions in this case. The result is that as we encounter the $t=0$ curvature singularity at $\tau \approx -0.467$, a cosmic jet develops with $\Gamma \to \infty$ and $(V_x, V_y, V_z) \to (0, 1, 0)$, since in this case $y$ is the direction of collapse when $t \to 0$.

\begin{figure}
\begin{center}
\includegraphics[width=5 cm]{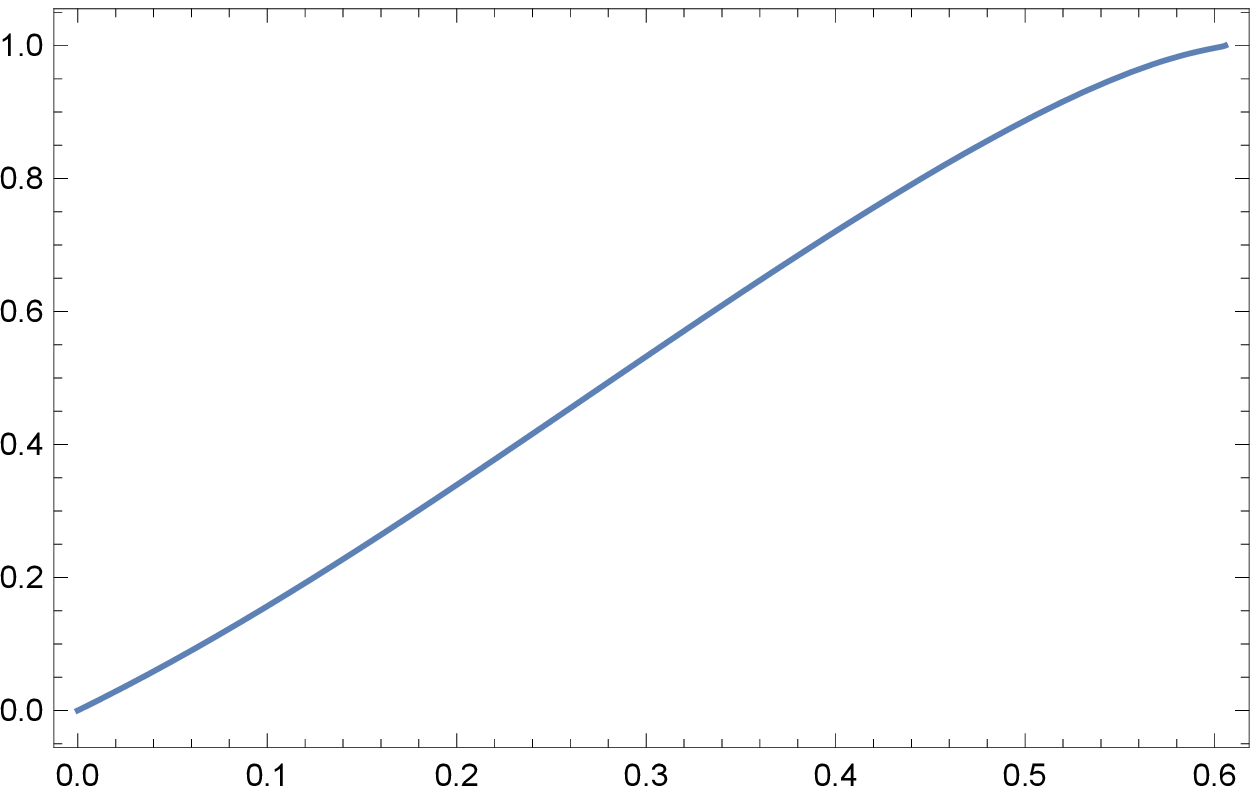}\quad \includegraphics[width=5 cm]{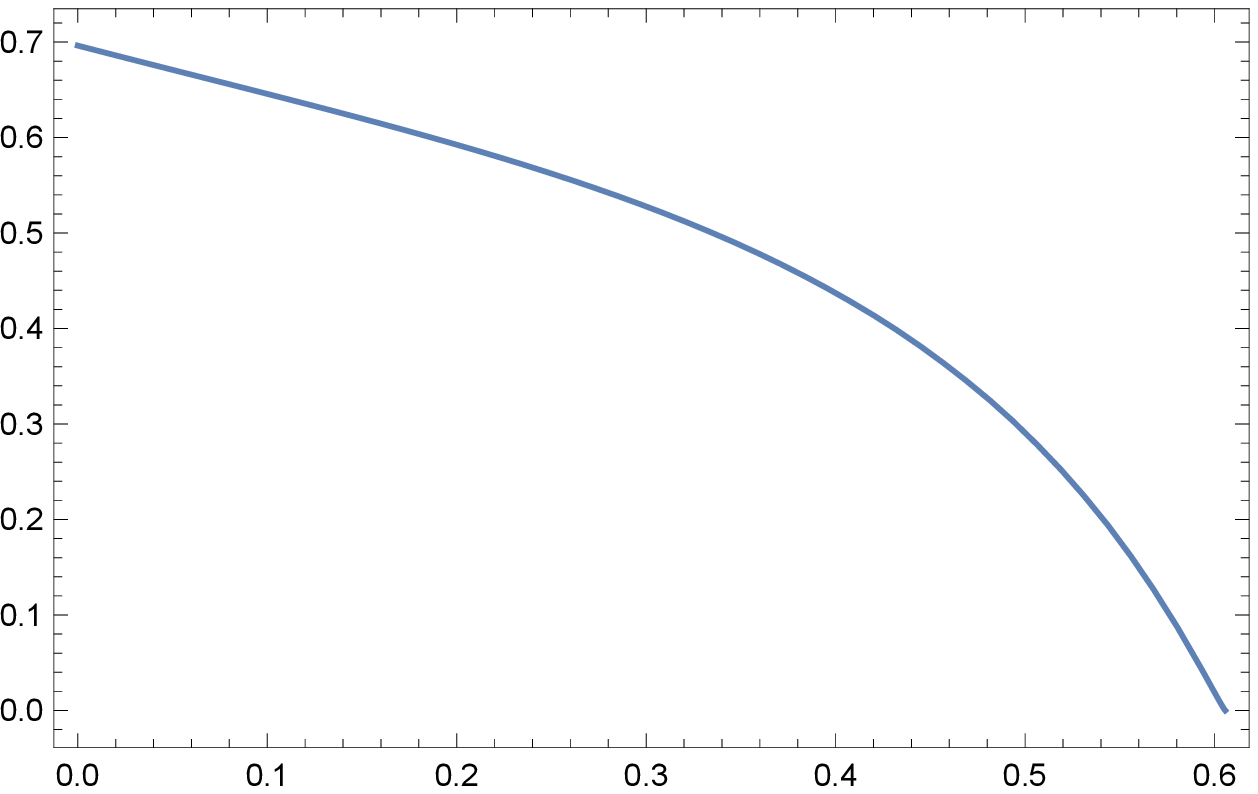}\quad \includegraphics[width=5 cm]{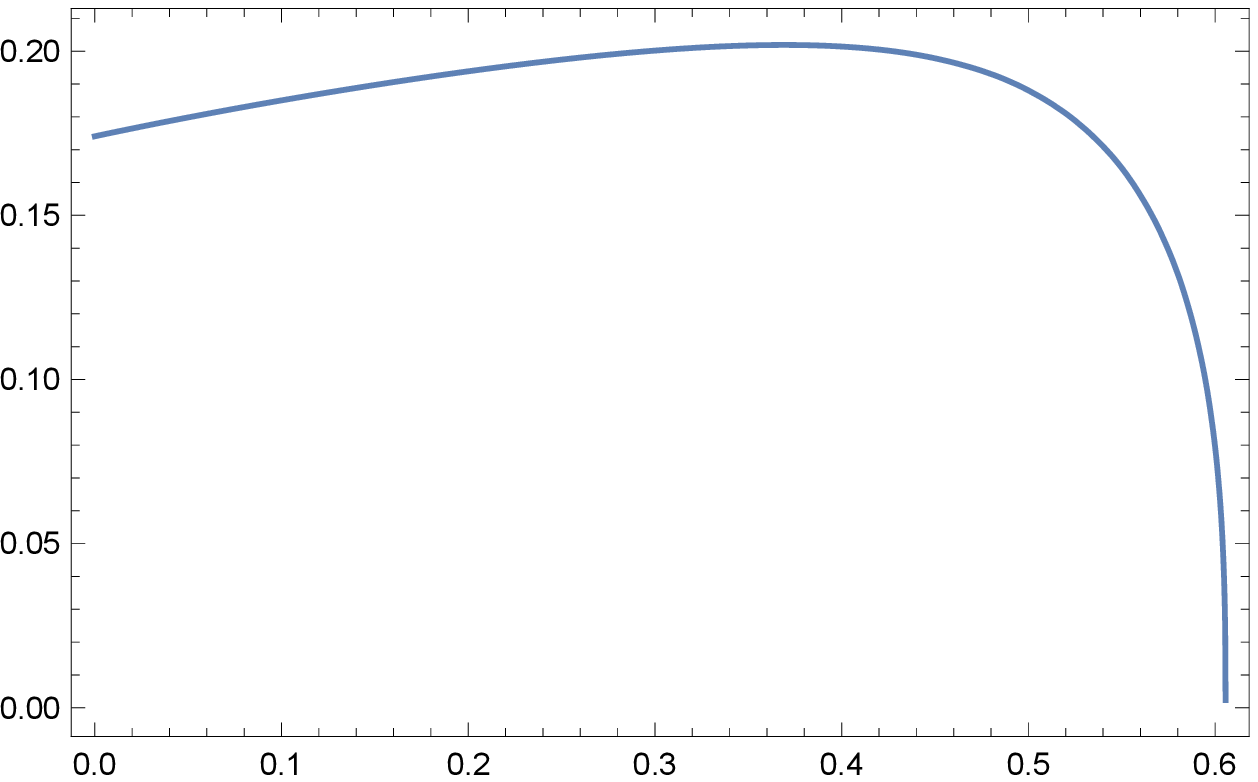}
\end{center}
\caption{ The figure depicts,  from left to right,  graphs of $V_x$, $V_y$ and $V_z$ versus proper time   $\tau$ for the parameter values $(\alpha, \beta) = (-1, -1)$ with $x(0) = 0.5$ at $\tau = 0$. Moreover, $t(0) = 0.5, \mathcal{W}(0) = 0, y(0) = z(0) = 0$ and $\mathcal{C}_y = \mathcal{C}_z = 1$.}
\label{fig3} 
\end{figure}

The  cosmic jets that  come about in the collapse scenario are all in principle double-jet configurations. It turns out that single cosmic jets also form in plane wave spacetimes~\cite{Bini:2014esa}. The wave scenario is further explored in sections V and VI.

\section{Wave Scenario: Plane Gravitational Wave}

The study of timelike geodesics of exact gravitational plane wave spacetimes has revealed that a single cosmic jet can asymptotically develop in the direction of motion of the plane wave~\cite{Bini:2014esa}. A more general treatment of this subject within the framework of linearized general relativity is contained in the recent work of Tucker and Walton~\cite{Tucker:2016wvt}. 

To illustrate the wave scenario, let us consider a spacetime with metric~\cite{R2}
\begin{equation}\label{L1}
 ds^2 = -2\,L^2\, du\,dv + u^{2s_2}\,dx^2 + u^{2s_3}\,dy^2\,,
\end{equation}
where $L$ is a constant length, $u$ and $v$ are related to the retarded and advanced null coordinates, respectively,  and 
\begin{equation}\label{L2}
s_2+s_3=s_2^2+s_3^2\,.
\end{equation}
Let us note that $s_2$ and $s_3$ cannot both be negative; either they are both positive, or one is positive and the other is negative. If either $s_2$ or $s_3$ is equal to zero or unity, this spacetime is flat. Otherwise, it can be shown that this metric belongs to a class of Petrov type N gravitational fields.  These spacetimes represent linearly polarized plane gravitational waves~\cite{BPR, JBG, Bini:2014esa}. A general discussion of pp waves is contained in section 24.5 of Ref.~\cite{R1}. Writing $u$ and $v$ in terms of standard coordinates $t$ and $z$, we have 
\begin{equation}\label{L3}
u := \frac{1}{\sqrt{2} L}\, (t-z)\,,  \qquad v := \frac{1}{\sqrt{2}L}\, (t+z)\,.
\end{equation}
In this section we measure all lengths in units of $L$; hence, we can in effect set $L=1$ for the rest of our considerations here.

As before, we can introduce observers that are at rest in this spacetime with natural tetrad frame
\begin{eqnarray}\label{L4}
e_{\hat 0} = \partial_t\,, \qquad
e_{\hat 1} = u^{-s_2}\,\partial_x\,,\qquad
e_{\hat 2} =  u^{-s_3}\,\partial_y\,, \qquad
e_{\hat 3} = \partial_z\,.
\end{eqnarray}
It turns out that these static observers are free and their tetrad frames are parallel propagated along their geodesic world lines. The spacetime curvature as measured by the static fiducial observers is given by
\begin{equation}\label{L5}
\mathcal{K}(u) := \frac{1}{2}\,(s_2-s_3)(s_2+s_3-1)\,\frac{1}{u^2}\,,
\end{equation}
see Appendix B. Thus there is a curvature singularity at the wave front $u=0$. We therefore limit our considerations to $u > 0$ for arbitrary $(s_2, s_3)$ subject to Eq.~\eqref{L2}. This spacetime admits five Killing vector fields, namely, 
\beq \label{L6}
\partial_v=(\partial_t + \partial_z)/\sqrt{2}\,, \qquad \partial_x\,, \qquad \partial_y\,
\eeq 
as well as
\beq \label{L7}
x\,\partial_v + \frac{u^{1-2s_2}}{1-2s_2}\, \partial_x\,, \qquad y\,\partial_v + \frac{u^{1-2s_3}}{1-2s_3}\,\partial_y\,
\eeq
for $s_2 \ne 1/2$ or $s_3 \ne 1/2$. Let us note that in case   $s_2 = 1/2$ and $s_3= (1\pm \sqrt{2})/2$, for instance, then instead of the Killing vector in Eq.~\eqref{L7} we have 
\beq \label{L8}
x\,\partial_v + \ln{u}\, \partial_x\,, 
\eeq
etc. In addition, the plane wave under consideration in this paper admits a homothetic vector field given by
\beq \label{L9}
 t\, \partial_t + (1-s_2)\,x\, \partial_x + (1-s_3)\,y\, \partial_y + z\, \partial_z\,.
\eeq

Let $ dx^\mu/d\tau = (\dot t, \dot x, \dot y, \dot z)$ denote the 4-velocity vector of free test particles in this spacetime. It follows from the existence of the null Killing vector  $\partial_v$ that 
\beq \label{L10}
 \dot t - \dot z = k_0\,,
\eeq
where $k_0 >0$ is a constant of the motion. Here, we have assumed that for a constant $z$, time increases with increasing proper time so that $\dot t >0$.  Moreover, we have 
\beq \label{L10a}
 \frac{du}{d\tau} = \frac{k_0}{\sqrt{2}}\,.
\eeq
Next, the spacelike Killing vectors
$\partial_x$ and $\partial_y$ imply that 
\beq \label{L11}
 \dot x = k_x\,u^{-2s_2}\,, \qquad   \dot y = k_y\,u^{-2s_3}\,,
\eeq
where $k_x$ and $k_y$ are constants of the motion as well. These results together with the fact that the 4-velocity vector is a timelike vector of unit length imply that 
\beq \label{L12}
 \dot t = \frac{1}{2} \left(k_0 + \frac{1}{k_0}\right) +\frac{1}{2k_0}\left(k_x^2\,u^{-2s_2}+ k_y^2\, u^{-2s_3}\right)\,,
\eeq
and 
\beq \label{L13}
 \dot z = \frac{1}{2} \left(-k_0 + \frac{1}{k_0}\right) +\frac{1}{2k_0}\left(k_x^2\,u^{-2s_2}+ k_y^2\, u^{-2s_3}\right)\,.
\eeq
Let us observe that with $k_0 =1$ and $k_x=k_y=0$, we recover the 4-velocity of the static fiducial observers. 

Projecting the 4-velocity of the free test particles onto the tetrad frame of the fiducial observers~\eqref{L4}, we find that relative to the static observers, the  4-velocity of the free particles $u^{\hat \alpha}$ is given by $\Gamma (1, V_x, V_y, V_z)$, where 
\beq \label{L14}
\Gamma =  \dot t\,, \qquad V_x = k_x\, \frac{u^{-s_2}}{\Gamma}\,, \qquad  V_y = k_y\, \frac{u^{-s_3}}{\Gamma}\,, \qquad V_z = \frac{\dot z}{\Gamma}\,.
\eeq
Inspection of these results indicate that two different situations can arise. If $s_2 >0$ and $s_3 >0$, then as $t \to \infty$, $\tau \to \infty$ and $u \to \infty$, so that 
\beq \label{L15}
 (V_x, V_y, V_z) \to (0, 0, V_0)\,, \qquad V_0 = \frac{1-k_0^2}{1+k_0^2}\,, \qquad \Gamma = \frac{1}{2} \left(k_0 + \frac{1}{k_0}\right)\,,
\eeq
which is a mild form of the cosmic jet. On the other hand, if $s_2$ or $s_3$ is negative, then 
\beq \label{L16}
 (V_x, V_y, V_z) \to (0, 0, 1)\,, \qquad \Gamma \to \infty\,.
\eeq
In either case, it is remarkable that the free test particles all line up parallel to the direction of motion of the plane wave. If $s_2 >0$ and $s_3 >0$, the proper distance along the $x$ and $y$ directions both \emph{expand} as time increases for a constant $z$, while if either $s_2$ or $s_3$ is negative, then one direction expands but the other contracts. It is interesting that the latter case leads to the formation of a cosmic jet for which $\Gamma \to \infty$; however, the direction of the cosmic jet is \emph{not} parallel to the direction of collapse; rather, it is in the direction of wave propagation.

\section{Nonplanar Wave Scenario}

Let us next consider a Ricci-flat solution of GR due to Harrison~\cite{Harrison:1959zz} that we write as
\begin{equation}\label{H1}
ds^2 = -\check{x}^{4/3}\, dt^2 + \lambda^2\, \check{u}^{6/5}\,dx^2+ \check{x}^{-2/3}\,\check{u}^{-2/5}\,dy^2 + \check{x}^{4/3}\,dz^2\,,
\end{equation}  
where
\begin{equation}\label{H2}
\check{x} := \frac{x}{T_0}\,, \qquad \check{u} := \frac{t-z}{T_0}\,.
\end{equation} 
Here, $T_0$ is a constant length and $\lambda > 0$ is a dimensionless parameter.  The dimensionless quantity $\check{u}$ is simply related to the retarded null coordinate $u$. Equation~\eqref{H1} can be obtained via straightforward coordinate transformations from the second degenerate solution obtained by Harrison and classified as  the ``III-2 ; $D_2$" metric~\cite{Harrison:1959zz}. This nonplanar gravitational wave spacetime is of type D in the Petrov classification~\cite{DIRC}.
The Kretschmann scalar for the Harrison solution can be expressed as 
\beq \label{H3}
 K = \frac{64}{27\,T_0^4\,\lambda^4}\, \check{x}^{-4}\,\check{u}^{-12/5}\,.
\eeq   
It follows that hypersurfaces  $x=0$ and the wave front $t-z=0$ are curvature singularities of the Harrison spacetime. We note that $\sqrt{-g}= \lambda\, |\check{x}|\,\check{u}^{2/5}$, which vanishes at these singularities. For the Harrison spacetime, 
\beq \label{H4}
\partial_t+ \partial_z\,, \qquad \partial_y\,
\eeq 
are null and spacelike Killing vector fields, respectively. In addition, there is a homothetic vector field given by
\beq \label{H5}
5\, t\, \partial_t + 6\,x\, \partial_x + 12\,y\, \partial_y + 5\, z\, \partial_z\,.
\eeq
The Harrison solution represents the propagation of nonplanar gravitational waves in the $z$ direction at the speed of light.

For the motion of free test particles in Harrison's spacetime, we find two constants of the motion from the projection of the 4-velocity of the free test particle on the null and spacelike Killing vector fields. That is, we have 
\begin{equation}\label{H6}
\dot t - \dot z =   \eta_0\, \check{x}^{-4/3}\,
\end{equation}
and 
\begin{equation}\label{H7}
\dot y = \eta_2 \, \check{x}^{2/3} \,\check{u}^{2/5}\,,
\end{equation}
where $\eta_0$ and $\eta_2$ are dimensionless constants. We assume $\eta_0 > 0$, so that for constant $z$, coordinate time increases monotonically  with proper time along the timelike geodesic world line.  It follows that 
\begin{equation}\label{H8}
\frac{d\check{u}}{d\tau} = \frac{\eta_0}{T_0}\, \check{x}^{-4/3}\,.
\end{equation}
Furthermore, $u_\mu\,u^\mu = -1$, implies that 
\begin{equation}\label{H9}
\eta_0\, (\dot t + \dot z)  =  1 + \lambda^2\, \dot {x}^2\, \check{u}^{6/5} +\eta_2^2 \,\check{x}^{2/3} \,\check{u}^{2/5}\,.
\end{equation}
The remaining geodesic equation for the motion of free test particles can be written as  
\begin{equation}\label{H10}
\lambda^2\, \frac{d}{d\tau}(\check{u}^{6/5}\, \dot x) = -\frac{2}{3x}\left[1+\lambda^2\, \dot {x}^2\, \check{u}^{6/5} + \frac{3}{2}\,\eta_2^2 \,\check{x}^{2/3} \,\check{u}^{2/5}\right]\,.
\end{equation}

It proves useful to define new quantities $\zeta$ and $W$,
\begin{equation}\label{H11}
\zeta := \check{u}^{1/5}\,, \qquad W := \check{u}^{1/5}\,\check{x}^{1/3}\,. 
\end{equation}
Then, using Eq.~\eqref{H8},  Eq.~\eqref{H10} can be expressed as the autonomous second-order differential equation
\begin{equation}\label{H12}
\frac{d^2W}{d\zeta^2} + h\, W^3 + h'\,W^5 = 0\,, 
\end{equation}
where $h$ and $h'$ are constants given by
\begin{equation}\label{H13}
h := \frac{50}{9}\,\frac{1}{\lambda^2\,\eta_0^2}\,, \qquad h' :=\frac{25}{3}\,\frac{\eta_2^2}{\lambda^2\,\eta_0^2}\,. 
\end{equation}
Integrating Eq.~\eqref{H12} once, we find 
\begin{equation}\label{H14}
\left(\frac{d W}{d \zeta} \right)^2  + \frac{h}{2}\, W^4 + \frac{h'}{3}\,W^6 = E\,, 
\end{equation}
where $E > 0$ is an integration constant. 

This result can be interpreted in terms of a one-dimensional motion of a particle with positive energy $E$ in a simple symmetric effective potential well; in fact,  the  motion is periodic with turning points $\pm W_0$,   where $W_0>0$ and $\pm W_0$ are the only real roots of 
\begin{equation}\label{H15}
E - \frac{h}{2}\, W^4 - \frac{h'}{3}\,W^6 = 0\,. 
\end{equation}
In the special case that $\eta_2=0$ and hence $h'=0$, $W(\zeta)$ can be expressed in terms of the Jacobi elliptic functions. 

It is important to recall here that the Kretschmann scalar $K$, apart from constant coefficients,  can be expressed as $W^{-12}$.   Thus, as $W$ periodically moves from $-W_0$ to $W_0$, only the half periods with  $W>0$ or $W<0$ are free of singularities  and such spacetime domains then occur between curvature singularities at $W=0$, where $dW/d\zeta = \pm \sqrt{E}$.  

To have a more explicit form of $W(\zeta)$, let us note that according to Eq.~\eqref{H11} if $\zeta=0$, $W=0$ for finite $x$. Henceforth, we assume that $W(0) = 0$. We can then express $W(\zeta)$ in the neighborhood of $\zeta=0$ as
\begin{equation}\label{H15a}
W(\zeta) = \pm \sqrt{E}\,\zeta\,\left[ 1 - \frac{h}{20}\,E\,\zeta^4 -\frac{h'}{42}\,E^2\,\zeta^6 +O(\zeta^8)\,\right]\,. 
\end{equation}
Let $2\,w$ be the period of the function $W(\zeta)$, which therefore vanishes at $\zeta_0=0$ and $\zeta_n = n\, w$, where $n=\pm 1, \pm 2, \pm3, \cdots$.

Let us turn now to the motion of free test particles relative to the fiducial observers at rest in this spacetime. The natural tetrad frame field of the static observers is given by 
\begin{eqnarray}\label{H16}
e_{\hat 0} = \check{x}^{-2/3}\,\partial_t\,, \qquad
e_{\hat 1} = \frac{1}{\lambda}\,\check{u}^{-3/5}\,\partial_x\,,\qquad
e_{\hat 2} =  \check{x}^{1/3}\,  \check{u}^{1/5}\,\partial_y\,, \qquad
e_{\hat 3} =  \check{x}^{-2/3}\, \partial_z\,.
\end{eqnarray}
Projecting $u^\mu = (\dot t, \dot x, \dot y, \dot z)$ on $e^{\mu}{}_{\hat \alpha}$ results in $u^{\hat \alpha} = \Gamma (1, V_x, V_y, V_z)$, where
\begin{equation}\label{H17}
\Gamma = \check{x}^{2/3} \dot t\,, \qquad V_x = \frac{\lambda}{\Gamma}\,\check{u}^{3/5} \dot x\,, \qquad V_y = \frac{\dot y}{W\,\Gamma}\,, \qquad  V_z = \frac{\dot z}{\dot t}\,.
\end{equation}
In terms of $\zeta$ and $W$,  $\check{x}^{2/3} = W^2/\zeta^2$ and $\check{u}^{3/5} = \zeta^3$; moreover, we can write the components of  $u^\mu = (\dot t, \dot x, \dot y, \dot z)$ as
\begin{equation}\label{H18}
\dot t = \frac{1}{2\eta_0} \left(1+  \eta_0^2\, \frac{\zeta^4}{W^4} + \eta_2^2\,W^2 + \frac{2}{h}\, \mathbb{W}^2\right)\,, 
\end{equation}
\begin{equation}\label{H19}
\dot x =  \frac{3\eta_0}{5\zeta^3}\, \mathbb{W}\,, 
\end{equation}
\begin{equation}\label{H20}
\dot y = \eta_2\, W^2\,, 
\end{equation}
\begin{equation}\label{H21}
\dot z = \frac{1}{2\eta_0} \left(1-  \eta_0^2\, \frac{\zeta^4}{W^4} + \eta_2^2\,W^2 + \frac{2}{h}\, \mathbb{W}^2\right)\,, 
\end{equation}
where
\begin{equation}\label{H22}
 \mathbb{W} = \frac{1}{W}\left(\frac{1}{W}\,\frac{dW}{d\zeta} - \frac{1}{\zeta}\right)\,. 
\end{equation}
Finally, we note that the Lorentz factor is given by
\begin{equation}\label{H23}
\Gamma = \frac{W^2}{2\eta_0\,\zeta^2} \left(1+  \eta_0^2\, \frac{\zeta^4}{W^4} + \eta_2^2\,W^2 + \frac{2}{h}\, \mathbb{W}^2\right)\,. 
\end{equation}

When free test particles approach the spacetime singularity at $W(\zeta)=0$, $\zeta$ is either $\zeta_0=0$ or $\zeta_n = n\, w$. The Lorentz factor $\Gamma$ is finite at $\zeta_0=0$, but diverges 
at $\zeta_n = n\, w$. Indeed, for $\zeta \to \zeta_n$, $dW/d\zeta \to \pm \sqrt{E}$ and a cosmic jet develops with
\begin{equation}\label{H24}
(V_x, V_y, V_z) \to  (\sigma, 0, \sigma')\,, 
\end{equation}
where $\sigma$ and $\sigma'$,  $\sigma^2 + \sigma'^2 =1$, can be simply computed using 
Eq.~\eqref{H17}. The result is
\begin{equation}\label{H25}
\sigma = \pm \frac{\eta_0\, \zeta_n^2\,\sqrt{2Eh}}{E+\frac{1}{2} \,\eta_0^2\, h \,\zeta_n^4}\,, \qquad
 \sigma'= \frac{E-\frac{1}{2}\, \eta_0^2\, h\, \zeta_n^4}{E+\frac{1}{2}\, \eta_0^2\, h\, \zeta_n^4}\,, 
\end{equation}
where the upper sign in $\sigma$ is for $|n| =$~even and the lower sign is for $|n| =$~odd.
The direction of jet motion is different from the $z$ axis, which is the direction of wave propagation. It is therefore clear that in this case the cosmic jet develops as the spacetime singularity at $\zeta_n$, $n = \pm 1, \pm 2, \pm3, \cdots$,  is approached.

\section{DISCUSSION}

We have searched for the formation of cosmic jets in certain simple exact solutions of GR that are mostly self-similar and hence admit homothetic Killing vector fields. In addition to the collapse and wave scenarios for cosmic jet formation, a new scenario has been identified in section VI in connection with a nonplanar gravitational wave spacetime and appears to be a new form of the wave scenario. 

In connection with the collapse scenario, let us consider a system that collapses under its own gravity such that it contracts along one spatial axis and expands along the other spatial axes. We are interested in the invariant velocity of the free test particles relative to the natural tetrad frame of the class of static observers in this gravitational field. We have demonstrated that in some simple exact spatially inhomogeneous perfect fluid solutions of GR, the magnitude of this relative velocity decreases along an expanding axis and asymptotically goes to zero as the expansion tends to infinity. On the other hand, the magnitude of the relative velocity increases along the contracting axis and asymptotically goes to the speed of light as the contraction tends to zero. This feature is observer independent and can be interpreted to mean that free test particles gain kinetic energy from the time-dependent gravitational field and asymptotically form a double-jet configuration. These conclusions are consistent with our previous results~\cite{Chicone:2010aa, Chicone:2010hy, Chicone:2010xr, Chicone:2011ie, Bini:2014esa}. The perfect fluid examples we have analyzed in this paper are much too simple and idealized to have any direct physical relevance;  however, it is clear from our numerical considerations that inhomogeneities can significantly interfere with the formation of cosmic jets. 

The specific type of spatial inhomogeneity that would \emph{not} impede the development of cosmic jets deserves further investigation. It appears from our numerical results that under reasonable physical conditions most inhomogeneities can block the tendency of anisotropically collapsing configurations toward cosmic jet formation. This circumstance may be related to the fact that relativistic jets are observed in only a small fraction of active galactic nuclei, while supermassive collapsed configurations are generally presumed to exist in the nuclei of most, if not all, massive galaxies. 

The formation of a cosmic jet is expected to be accompanied by a strong burst of electromagnetic radiation, since some of the test particles in the ambient medium are presumed to be electrically charged.

\appendix

\section{Mars-Wolf Metric}

The Mars-Wolf solution under consideration in this paper is in fact Solution A given in Ref.~\cite{Mars:2002pn}, namely, 
\begin{equation}\label{A1}
ds^2=t^\alpha\, X^{-2+2\alpha^2/q}\,(-dt^2 + dX^2 + t^{1-\alpha-\beta}\, dy^2 + t^{1-\alpha +\beta}\,dz^2)\,,
\end{equation}
where
\begin{equation}\label{A2}
q= \beta^2 + \alpha^2 -2\alpha -1 \ne 0\,.
\end{equation}
A discussion of this solution is contained on page 569 of Ref.~\cite{R1}. In its general form, this metric is of type I in the Petrov classification. 

For $\alpha=0$, let $X= \exp{x}$ and $\beta = -b$; then, Eq.~\eqref{A1} reduces to the Petrov type D metric~\eqref{M1} discussed in section III. Henceforth, we assume that $\alpha \ne 0$. 

If $q=\alpha^2$, the Mars-Wolf solution reduces to the Kasner solution. We thus define
\begin{equation}\label{A3}
Q := 1- \frac{q}{\alpha^2} = \frac{2\alpha +1 -\beta^2}{\alpha^2}\,
\end{equation}
and introduce Kasner-like parameters $p_1=\alpha/(\alpha +2)$, $p_2 = (1-\beta)/(\alpha +2)$ and $p_3 = (1+\beta)/(\alpha +2)$. With simple coordinate transformations involving $t$ and $X$, the Mars-Wolf metric takes the form~\eqref{W1} in section IV. We note that in this form the metric is conformally related to a Kasner-like metric for which $p_1+p_2+p_3 =1$, but $p_1^2 + p_2^2 + p_3^2 =1-2Qp_1^2$.

\section{Curvature of the Plane Wave}  

The components of the Riemann curvature tensor of the plane wave discussed in section V as measured by the static reference observers are given by
\begin{equation}\label{B1}
R_{\hat \alpha \hat \beta \hat \gamma \hat \delta} = R_{\mu \nu \rho \sigma}\, e^{\mu}{}_{\hat \alpha}\,e^{\nu}{}_{\hat \beta}\, e^{\rho}{}_{\hat \gamma}\,e^{\sigma}{}_{\hat \delta}\,,
\end{equation}
where the tetrad frame $e^{\mu}{}_{\hat \alpha}$ is given by Eq.~\eqref{L4}. 
One can express these quantities as a $6\times 6$ matrix $(R_{IJ})$, where $I$ and $J$ are indices that belong to the set $\{01,02,03,23,31,12\}$. In a Ricci-flat spacetime, $(R_{IJ})$ can be written as 
\beq
\label{B2}
\left[
\begin{array}{cc}
{\mathcal E} & {\mathcal B}\cr
{\mathcal B} & -{\mathcal E}\cr
\end{array}
\right]\,,
\eeq
where ${\mathcal E}$  and ${\mathcal B}$ are symmetric and traceless $3\times 3$ matrices. For the plane wave under consideration in section V, we find 
\beq
\label{B3}
{\mathcal E}={\mathcal K}(u)\left[
\begin{array}{ccc}
-1&0 & 0\cr
0&1 & 0\cr
0&0 & 0\cr
\end{array}
\right]\,,\qquad
{\mathcal B}={\mathcal K}(u)\left[
\begin{array}{ccc}
0&1 & 0\cr
1&0 & 0\cr
0&0 & 0\cr
\end{array}
\right]\,,
\eeq
where ${\mathcal K}(u)$, given by Eq.~\eqref{L5}, can also be written as
\begin{equation}\label{B3a}
\mathcal{K}(u) = \frac{s_2 (s_2-1)}{u^2} = -\frac{s_3 (s_3-1)}{u^2}\,.
\end{equation}
It follows that if either $s_2$ or $s_3$ vanishes, then the spacetime is flat. 

In connection with ${\mathcal E}$  and ${\mathcal B}$, the ``gravitoelectric" and ``gravitomagnetic" components of the Weyl tensor, respectively, an error in Eq.\ (25) of Ref.~\cite{Bini:2014esa} must be corrected: The $3 \times 3$ matrices must be the same as in our Eq.~\eqref{B3}.

The metric of the plane wave in section V depends upon $s_2$ and $s_3$ such that $s_2 +s_3 = s_2^2 + s_3^2$. It follows that if $s_2 < 0$, then $s_3^2-s_3 = s_2 -s_2^2 <0$ and hence $0 < s_3 < 1$. By symmetry, if $s_3 < 0$, then $0 < s_2 < 1$. The relationship between $s_2$ and $s_3$ can be written in terms of an angular parameter $\theta$,  $ 0\le \theta < 2\,\pi$, as 
\begin{equation}\label{B4}
s_2 = \frac{1}{2} (1 + \sqrt{2}\, \cos \theta)\,, \qquad s_3 = \frac{1}{2} (1 + \sqrt{2}\, \sin \theta)\,.
\end{equation}
We note that for $3\,\pi/4 < \theta < 5\,\pi/4$, we have $s_2 <0$ and $s_3>0$, while for  
 $5\,\pi/4 < \theta < 7\,\pi/4$, we have $s_2 >0$ and $s_3<0$. Otherwise, $s_2 \ge 0$ and $s_3\ge 0$. 

In terms of $\theta$, ${\mathcal K}(u)$ can be written as 
\begin{equation}\label{B5}
{\mathcal K}(u) =  \frac{1}{4u^2}\, \cos {2 \theta}\,.
\end{equation}
We recall that the spacetime curvature vanishes if either $s_2$ or $s_3$ is zero. To illustrate this point explicitly for $s_3 = 0$ and $s_2 = 1$, consider metric~\eqref{L1} in this case, namely,   
\begin{equation}\label{B6}
ds^2 =  -2\, du\, dv + u^2\,dx^2 + dy^2\,
\end{equation}
and let $(v, x) \mapsto (\xi, \zeta)$, where
\begin{equation}\label{B7}
v = \xi - \frac{1}{2} \frac{\zeta^2}{u}\,, \qquad x = \frac{\zeta}{u}\,.
\end{equation}
Then, 
\begin{equation}\label{B8}
  -2\, du\, dv + u^2\,dx^2 = -2\,du\,d\xi + d\zeta^2\,. 
\end{equation}
Thus the plane wave spacetime reduces to Minkowski spacetime if either $s_2$ or $s_3$ vanishes as a consequence of the symmetry between $s_2$ and $s_3$.

\section*{Acknowledgments}
D.B. thanks the Italian INFN (Salerno) for partial support.

\end{document}